\documentclass[aps,twocolumn,superscriptaddress,prl,showpacs]{revtex4}

\usepackage{graphicx}

\begin{document}
\title{Dynamic nuclear polarization of a single charge-tunable  InAs/GaAs quantum dot}


\author{B.~Eble}
\author{O.~Krebs}
\email[Corresponding author : ]{Olivier.Krebs@lpn.cnrs.fr}
\author{A.~Lema\^{i}tre}
\affiliation{CNRS-Laboratoire de Photonique et Nanostructures, Route de Nozay, 91460
Marcoussis, France}
\author{K.~Kowalik}
 \affiliation{CNRS-Laboratoire de Photonique et Nanostructures, Route de Nozay, 91460
Marcoussis, France}\affiliation{Institute of Experimental Physics, Warsaw University, Ho\.{z}a
69, 00-681 Warsaw, Poland}
\author{A.~Kudelski}
\author{P.~Voisin }
\affiliation{CNRS-Laboratoire de Photonique et Nanostructures, Route de Nozay, 91460
Marcoussis, France}
\author{ B. Urbaszek}
\author{ X. Marie}
\author{ T. Amand}
\affiliation{Laboratoire de  Nanophysique Magn\'{e}tisme et Opto\'{e}lectronique, INSA, 31077 Toulouse
Cedex 4, France}
\def\xp{$X^{+}$ }
\def\xm{$X^{-}$ }
\def\x0{$X^{0}$ }


\date{July 12, 2005}

\begin{abstract}
We report on  the dynamic nuclear polarization of a single charge-tunable
self-assembled InAs/GaAs quantum dot in a longitudinal magnetic field of
$\sim$0.2T. The hyperfine interaction between the optically oriented electron
and nuclei spins leads to the polarization of the quantum dot nuclei measured
by the Overhauser-shift of the singly-charged excitons ($X^{+}$ and $X^{-}$).
When going from $X^{+}$ to $X^{-}$, we observe a  reversal of this  shift which
reflects the average electron spin optically written down in the quantum dot
either in the    $X^{+}$ state or in the final state of $X^{-}$ recombination.
We discuss a theoretical model which indicates an efficient depolarization
mechanism for the nuclei limiting their polarization to $\sim$10\%.
\end{abstract}

\maketitle

\begin{figure}[h]
\includegraphics[width=0.48 \textwidth,keepaspectratio]{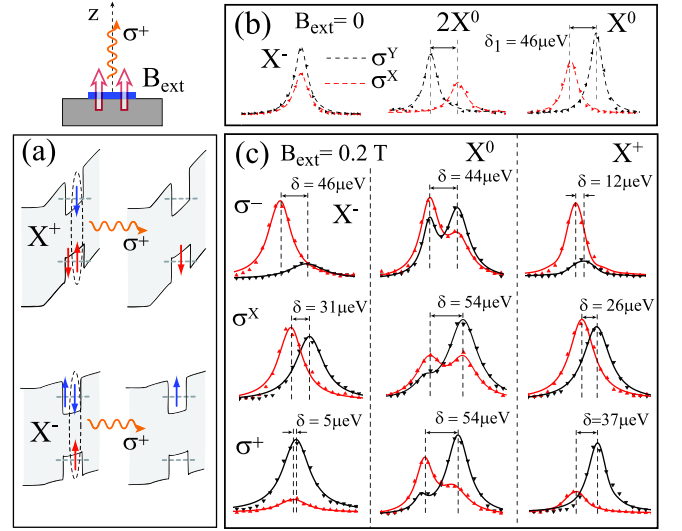}
\caption{(Color online) (a) Sketch of the spin configuration for $X^{+}$ and $X^{-}$ for
$\sigma^{+}$ emission. (b) and (c) Zoom over a 200~$\mu$eV range of PL spectra excited at
1.34~eV in zero and 0.2~T magnetic field respectively.  (b) Detection  in linear polarization.
(c) Detection in $\sigma^{-}$ (red) and  $\sigma^{+}$ (dark) for three different excitation
polarizations as indicated on the left.}
 \label{Fine-Structure}
\end{figure}
\begin{figure}[h]
 \includegraphics[width=0.48\textwidth,keepaspectratio]{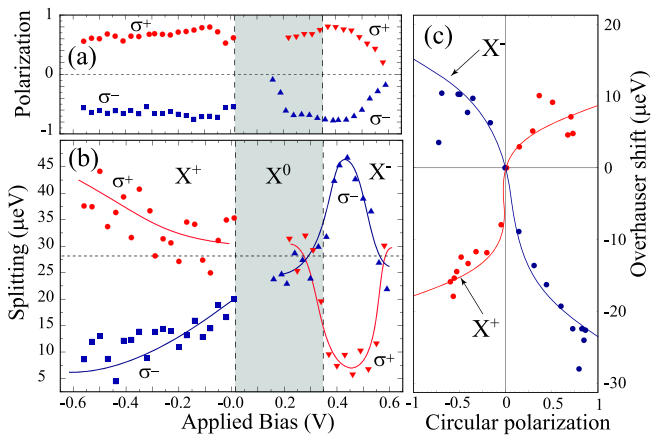}
 \caption{(Color online)(a) circular polarization and (b) spin splitting   of $X^{+}$ and
$X^{-}$ PL lines   against  applied bias  at $B_{ext}=$0.2~T,   under $\sigma^{+}$ (red) and
$\sigma^{-}$ (blue) polarized excitation  at 1.31~eV. The gray-shaded area represents the
region of $X^{0}$ stability and the solid lines are a guide for the eye. (c) Overhauser shift
versus circular polarization  for $X^{-}$ (+0.4~V) and $X^{+}$ (-0.2~V) measured for
excitation polarizations varying from $\sigma^{+}$ to $\sigma^{-}$. Solid lines are
theoretical fits according to Eq.~\ref{OH} obtained with
$\frac{g_{e}}{g_{X}}\delta_{Z}=$~7~$\mu$eV, $\Delta^{\star}$=1.3~(1.3)~meV,
$\tau_{c}$=0.6~(0.06)~ns and $\kappa^{-1/2}$=1.8~(3.8)~$\mu$eV for $X^{+}$($X^{-}$)
respectively.}
 \label{OS&PL_vs_bias}
 \end{figure}

\indent Spin dynamics of an electron confined in a self-assembled semiconductor quantum dot
(QD) is currently the subject of an intense research
\cite{Nat-Kroutvar,PRL94-Braun,PRB65-Merkulov2002,PRL88-Khaetskii-Loss,PRL94-Smith,PRL94-Laurent,PRL94-Gurudev}.
It indeed represents a promising direction  for implementing quantum computation algorithms in
solid state, because once the electron is confined in a quantum dot, its spin dynamics at low
temperature is \textit{almost} no longer subjected to the random perturbations which lead to
relaxation and decoherence in bulk or quantum wells. For example the usual spin relaxation due
to spin-orbit interaction turns out to be quite negligible
 \cite{PRB61-Khaetskii,Nat-Kroutvar}. The subsisting sources of relaxation which  have been
identified in real QDs are (i) the exchange interaction with additional hole(s) or electron(s)
\cite{PRL94-Laurent} and (ii) the hyperfine interaction with the QD nuclei spins
 \cite{PRB65-Merkulov2002,PRL88-Khaetskii-Loss}. In order to address this issue with optical
techniques,  field-effect structures embedding charge-tunable QDs \cite{PRL79-Warburton} offer
an amazing potential : the exchange-induced spin relaxation can be kept under control e.g. by
extracting the hole from a photo-excited electron-hole pair (named further exciton)
\cite{Nat-Kroutvar} or by adding a charge preventing the exchange to operate
\cite{PRL94-Laurent,PRB68-Flissikowski}, while the effective role of the hyperfine interaction
which only affects the conduction band electrons can be investigated  by controlling
the nature (electron or hole) of the spin-polarized  carrier.\\
\indent In this Letter we  address the issue of hyperfine interaction in a self-assembled
charge-tunable InAs/GaAs quantum dot submitted to  an  external magnetic field of $\sim$0.2~T.
Optical excitation with circularly polarized light is used for writing down the electron
and/or hole spins, whereas an external bias applied to the n-Schottky-type sample controls the
electronic charge~\cite{PRB63-Finley-2,PRL94-Laurent,PRL79-Warburton,APL86-Ediger}. The same
single QD has been studied in three different regimes:   when the   electron spin
$\hat{S}^{e}$ interacting with the nuclear spins $\hat{I}^{j}$ (i) forms with two
photo-excited holes a positive trion $X^{+}$, (ii)  forms a neutral exciton $X^{0}$, and (iii)
results from the radiative recombination of a negative trion $X^{-}$ made of two electrons and
one hole. By measuring the Overhauser shift  of the $X^{+}$ or $X^{-}$ Zeeman splitting
\cite{PRL86-Gammon-Merku,PRB71-Yokoi,PRL94-Bracker}, we show that the small applied magnetic
field leads to the optically-induced polarization of the QD nuclei, with a non-linear
dependence on the average electron spin $\langle \hat{S}^{e}_{z}\rangle$ deduced from  the
photoluminescence (PL) circular polarization. Remarkably, this shift changes sign with the
crossing from $X^{+}$ to $X^{-}$. We present a theoretical description of the nuclear
polarization dynamics  in InAs/GaAs
QDs explaining most of our results.\\
\indent The sample which has already been used in Ref.~\cite{PRL94-Laurent} was grown by
molecular beam epitaxy on a [001]-oriented semi-insulating GaAs substrate.  The InAs QDs are
grown in the Stranski-Krastanov mode  25~nm above a 200nm-thick n$^{+}$-GaAs layer and capped
by an intrinsic GaAs~(25~nm)/Al$_{0.3}$Ga$_{0.7}$As~(120~nm)/GaAs~(5~nm) multilayer. The QD
charge is controlled by an electrical bias  applied between a top Schottky contact  and a back
ohmic contact. We used a metallic mask evaporated on the Schottky gate with 1$\mu$m-diameter
optical apertures to  spatially select single QDs.\begin{figure}[h]
\includegraphics[width=0.48 \textwidth,angle=0]{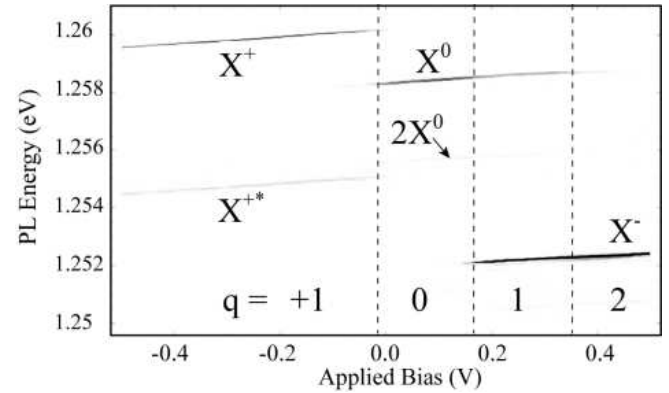}
\caption{Gray-scale contour plot of the PL intensity from a single InAs QD at T=5~K versus the
detection energy and applied bias under  intra-dot excitation at 1.31~eV.} \label{PL-map}
\end{figure}\\
\indent Figure \ref{PL-map} shows the T=5K  PL intensity contour plot against bias and
detection energy of the single QD that has been extensively studied in this work.  The
identification of the different spectral lines and of the associated QD charge relies on
several robust observations : (i) Between 0~V and $\sim$0.15~V we only observe the PL emission
from the  ground state exciton $X^{0}$  clearly identified by its fine structure
[Fig.~\ref{Fine-Structure}(b)] \cite{PRB65-Bayer} and by the biexciton (hardly perceptible in
Fig.~\ref{PL-map}) appearing under stronger excitation at lower energy. (ii) Above 0.15~V the
$X^{-}$ trion red-shifted by $\sim$6~meV shows up indicating the charging of the QD with an
electron~\cite{PRL90-Urbaszek,PRB63-Finley,PRL94-Laurent}. Both lines ($X^{0}$ and $X^{-}$)
still coexist because under non strictly-resonant excitation (here 1.31~eV) a single
photo-hole can be created in  the QD giving rise to optical recombination with the resident
electron. (iii) Above 0.35~V, the neutral exciton line definitely disappears indicating the
occupation with 2 electrons. (iv) For negative bias a symmetrical charging effect occurs for
 holes. The neutral exciton line which disappears as a result of the electron tunnelling
out of the QD, is replaced by a 3~meV blue-shifted line  assigned to the trion $X^{+}$
\cite{APL86-Ediger,XplusStar}. Although the applied bias only controls the conduction band
chemical potential and thus cannot itself generates the QD charging with holes, this effect is
achieved under  strong intra-dot excitation. It directly creates a hole within the QD, which
does not escape as the electron thanks to its larger effective mass.\\
 \indent To study the influence of hyperfine interaction on spin dynamics, optical orientation
experiments have been performed  in presence of a small  longitudinal magnetic field parallel
to the QD growth axis $z$~\cite{Optical-Orientation,PRL86-Gammon-Merku,PRL94-Braun}. The
latter was provided  by a permanent magnet simply put below the sample within the cryostat
cold finger. Its amplitude $B_{ext}$ at the sample position was estimated to $\approx$0.2~T.
We used a  standard micro-PL setup based on a $\times 50$~microscope objective, a double
 spectrometer of 0.6m-focal length and a Nitrogen-cooled CCD array detector, providing a spectral
resolution of 30~$\mu$eV and a precision   on line position of about 1~$\mu$eV after
deconvolution by a Lorentzian fit. The optical excitation and detection were both performed
along the $z$ axis. Thus the degree of PL circular polarization defined by
$\rho_{c}=(I_{\sigma^{+}}-I_{\sigma^{-}})/(I_{\sigma^{+}}+I_{\sigma^{-}})$, where
$I_{\sigma^{+(-)}}$ denotes the PL intensity  measured in $\sigma^{+(-)}$ polarization,
  traces the average  spin $\langle S^{e}_{z}\rangle =-\rho_{c}/2$ of the electron
participating in the PL signal \cite{Optical-Orientation}. This results from the usually
accepted assumption of pure heavy-hole  ground state with angular momentum projection
$m_{z}=\pm\frac{3}{2}$ in InAs QDs leading to  optically active electron-hole pairs
$|\pm1\rangle=|\mp\frac{1}{2},\pm\frac{3}{2}\rangle$. As a result, depending on the QD charge
state, we are able to read out either  the spin of the electron for both $X^{+}$ and $X^{0}$
states, or  the spin of the hole in $X^{-}$ from which we deduce the electron spin   $\langle
S^{e}_{z}\rangle =+\rho_{c}/2$ left in the QD after the optical recombination as sketched in Fig.~\ref{Fine-Structure}(a).\\
 Figure  \ref{Fine-Structure} shows the influence of the  applied magnetic field on the
(charged) exciton fine structure. In InAs QDs, the electron-hole exchange interaction leads to
the splitting of  $X^{0}$  between the dark states (angular quantum number $J$=2) uncoupled to
light and the bright states ($J$=1) which form the $X^{0}$ line observed in PL experiments.
Due to  in-plane  anisotropy of real QDs the latter is further split by an energy
$\delta_{1}$$\sim$50~meV into $|X\rangle$ and $|Y\rangle$ eigenstates linearly polarized along
the crystallographic axes $\langle 110\rangle$ \cite{PRB67-Bester-FS,PRB65-Bayer}. This is
illustrated in Fig.~\ref{Fine-Structure}(b) which presents linear polarization resolved
spectra in zero magnetic field. The $X^{0}$ line is split by $\delta_{1}$=46$\mu$eV  and the
biexciton  $2X^{0}$ shows the same splitting with reversed sequence of polarization
\cite{PRL85-Besombes}. In contrast, the trion lines  $X^{+}$ and $X^{-}$ are not split  in
agreement with the vanishing of the exchange interaction between one electron (hole) and two
holes (electrons) forming a spin singlet in the ground state~\cite{APL81-Akimov}. In magnetic
field, the Zeeman interaction separates the $\sigma^{\pm}$-polarized components of the trion lines
by $\delta_{Z}=|g_{X}|\mu_{B}B_{ext}$ where $g_{X}$ is the exciton $g$-factor (supposed to be
constant for the three lines considered here) and $\mu_{B}$=58$\mu$eV/T is the Bohr magneton.
It also increases the bright $X^{0}$ splitting to $\sqrt{\delta_{1}^{2}+\delta_{Z}^{2}}$.
Figure \ref{Fine-Structure}(c) shows the PL spectra resolved in  circular polarization.  Under
linearly  polarized excitation we find for the three lines a  Zeeman splitting
$\delta_{Z}\approx$28~$\mu$eV in agreement with an exciton $g$-factor of
$\sim$3~\cite{PRB65-Bayer}. Under circularly polarized excitation a significant deviation from
the sole Zeeman interaction is now observed :  the $X^{+}$ splitting gets larger in
$\sigma^{+}$ excitation by +10~$\mu$eV and smaller in $\sigma^{-}$ by -15~$\mu$eV. This
so-called Overhauser shift (OHS) denoted  $\delta_{n}$ indicates the polarization of the QD
nuclear spins which progressively builds up through the hyperfine interaction with the
optically oriented electrons in the QD. Remarkably a symmetrical but reversed effect occurs
for $X^{-}$ with a  shift $\delta_{n}$=+15~$\mu$eV in $\sigma^{-}$ and
$\delta_{n}$=-25~$\mu$eV in $\sigma^{+} $, whereas  the PL from $X^{-}$ and $X^{+}$ shows the
same helicity. This OHS reversal  demonstrates that in the case of a spin-polarized $X^{-}$
for which the total electron spin is zero, the mechanism leading to nuclear polarization
doesn't operate during  $X^{-}$ lifetime but takes place due to the interaction  with the
single electron left in the QD after the optical recombination. This contrasts with the
results reported for GaAs QDs~\cite{PRL94-Bracker}.  For $X^{0}$ which still shows a weak
polarization at 0.2~T, we observed no significative OHS except when there is an overlapp with
$X^{+}$ (situation of Fig.~\ref{Fine-Structure}(c) because of
   excitation at higher energy), in which case it only acts as a probe of OHS produced by $X^{+}$.
An other  feature of these results is the pronounced OHS asymmetry when changing the
excitation from $\sigma^{+}$ to $\sigma^{-}$. This clearly appears in Fig.~\ref{OS&PL_vs_bias}
which reports the bias dependence of circular polarization and of the trion spin splitting.
This asymmetry means that polarizing the nuclear spins in the direction which produces a
larger effective field for the electron  is more difficult than in the opposite direction. The
total electron spin splitting represents indeed the main energy cost of the electron-nuclei
\textit{flip-flop} process (the Zeeman splitting of nuclei being much smaller), which thus
produces a negative feedback on the nuclear polarization as we show in the following.
\\
 \indent The Hamiltonian describing the hyperfine interaction of a  single  electron spin
$\hat{S}^{e}=\frac{1}{2}\hat{\sigma}^{e}$ confined in a QD with $N$ nuclear spins is given by
 \cite{Optical-Orientation,PRL86-Gammon-Merku}:
\begin{equation}\label{H-hyperfin}
\hat{H}_{hf}=\frac{\nu_{0}}{2}\sum_{j}A^{j}\left|\psi(\mathbf{r}_{j})\right|^{2}\left(\hat{I}_{z}^{j}\hat{\sigma}^{e}_{z}+\frac{\hat{I}_{+}^{j}\hat{\sigma}^{e}_{-}+\hat{I}_{-}^{j}\hat{\sigma}^{e}_{+}}{2}
\right)
\end{equation}
where $\nu_{0}$ is the two-atom unit cell volume, $\mathbf{r}_{j}$ is the position of the
nuclei $j$ with spin $\hat{I}^{j}$, $A^{j}$ is the constant of hyperfine interaction with the
electron and $\psi(\mathbf{r})$ is the electron envelope function. The sum goes over the
nuclei interacting significantly with the electron (i.e. essentially in the effective QD
volume defined by $V= (\int \left|\psi(\mathbf{r})\right|^{4}d\mathbf{r})^{-1}=\nu_{0}N/2$).
This interaction has two important effects on the electron-nuclei spin system. (i) It  acts as
an effective magnetic field $\mathbf{B}_{n}\approx\sum_{j}A^{j}\mathbf{I}^{j}/(N
g_{e}\mu_{B})$ on the electron spin of  $g$-factor $g_{e}$. In absence of nuclear
polarization, this random nuclear field  averages to zero but shows fluctuations $\propto
A/\sqrt{N}$ of the order of 30~mT\cite{PRB65-Merkulov2002,PRL94-Braun}. In a classical
description the electron spin precess around the total magnetic field
$\mathbf{B}=\mathbf{B}_{ext.}+\mathbf{B}_{n}$, which determines the spin dynamics of a single
electron\cite{PRB65-Merkulov2002} as well as of $X^{+}$ \cite{PRL94-Braun}. (ii) Since this
precession stops randomly within a correlation time $\tau_{c}$ (due to optical
excitation/recombination or QD charging), the conservation of angular momentum leads to the
transfer of spin polarization towards the nuclei. Quantum mechanically this flip-flop
mechanism corresponds to the second term of Eq.~(\ref{H-hyperfin}). The nuclear polarization
can then accumulate in the QD giving rise to the OHS through the first term of
Eq.~(\ref{H-hyperfin}) under the condition that the spin diffusion driven by the dipole-dipole
interaction between nuclei  is quenched, which is in principle largely achieved for fields
above 1~mT~\cite{Optical-Orientation}.\\ \indent To derive from Eq.~(\ref{H-hyperfin}) a
convenient expression for the nuclear polarization dynamics we assume as a first order a
uniform electron wavefunction $\psi(\mathbf{r})=\sqrt{2/N\nu_{0}}$ over the involved nuclei.
This also amounts to consider a uniform nuclear polarization $\rho_{\alpha}=\langle
I_{z}^{\alpha} \rangle/I^{\alpha}$ for each isotopic specie  $\alpha$  of the QD. The theory
of time-dependent perturbation up to the second order applied to the flip-flop interaction
characterized by a correlation time $\tau_{c}$ yields the following equation rate
\cite{Abragam}:
\begin{equation}\label{DNP}
\frac{d\langle I^{\alpha}_{z}\rangle}{dt}=-\frac{1}{T^{\alpha}}\left(\langle
I^{\alpha}_{z}\rangle-Q^{\alpha}\langle S_{z}^{e}\rangle\right)
\end{equation}
where $Q^{\alpha}=\frac{I^{\alpha}(I^{\alpha}+1)}{S(S+1)}$ and  $T^{\alpha}$ is given by:
\begin{equation}\label{time}
\frac{1}{T^{\alpha}}=\frac{2f_{e}\tau_{c}}{1+\left[(\frac{g_{e}}{g_{X}}\delta_{Z}+\delta_{n})\tau_{c}/\hbar\right]^{2}}\left(\frac{A^{\alpha}}{N\hbar}\right)^{2}
\end{equation}
where $f_{e}$ is the  fraction of time that the QD contains an electron, which obviously
depends on the excitation power. The OHS is then related to the average nuclear spins by
$\delta_{n}=-2\sum_{\alpha}x_{\alpha}\langle I_{z}^{\alpha}\rangle A^{\alpha}$ where $
x_{\alpha}$ is the fraction of specie $\alpha$. Note Eq.~(\ref{DNP}) is only valid under the
condition of weak nuclear polarization~\cite{Abragam,Optical-Orientation}. This is
experimentally verified here with a maximum polarization $\rho_{\alpha}\approx 0.1$ deduced
from the measured OHS ($\approx$25~$\mu$eV) divided by  its  maximum theoretical  value
($\approx$250$\mu$eV for a realistic InGaAs QD). The stationary solution of Eq.~(\ref{DNP})
driven by the electron polarization $\langle S_{z}^{e}\rangle/S$ close to unity is therefore
far from being reached which means that nuclear depolarization must be taken into account. The
physical origin of this mechanism likely relies on the dipolar (or quadrupolar) coupling
between nuclei, that in spite of  the screening by the applied magnetic field opens a way for
nuclear spin diffusion due to the time-dependent hyperfine interaction with  electron
spin\cite{PRL86-Gammon-Merku}. However, since we could not investigate further this effect by
varying the field, we simply describe it by adding to Eq.~(\ref{DNP}) the term $-f_{e}\langle
I^{\alpha}_{z}\rangle/T_{d}$ where $T_{d}$ is a depolarization time constant independent on
$\alpha$. The OHS reached at equilibrium is then given by the implicit equation:
\begin{equation}\label{OH}
\delta_{n}=\frac{-\Delta^{\ast}\langle S_{z}^{e}\rangle}{1+\kappa\left[
(\hbar/\tau_{c})^{2}+(\frac{g_{e}}{g_{X}}\delta_{Z}+\delta_{n})^{2}\right]}
\end{equation}
where $\Delta^{\ast}=2\tilde{A}\sum_{\alpha}x_{\alpha}Q^{\alpha}$,
$\kappa=\tau_{c}/T_{d}(N/\tilde{A})^{2}$ and we have used $\tilde{A}$ ($\approx$50~$\mu$eV)
instead of  $A^{\alpha}$ which indeed weakly depends on $\alpha$. Here we treat
$\Delta^{\ast}$, $\tau_{c}$ and $\kappa$ as fitting parameters, while $\delta_{n}$,
$\delta_{Z}$, $\langle S_{z}^{e}\rangle$  can be determined from the experiments. Note yet
that $\Delta^{\ast}$  amounts to  $\approx$1.3~meV  for a realistic In(Ga)As QD (with
$x_{In}=0.3$, $x_{Ga}=0.2$, $x_{As}=0.5$).\\\indent Equation (\ref{OH})  clearly shows  the
negative feedback of the OHS on its equilibrium value through the electron spin splitting
$(\frac{g_{e}}{g_{X}}\delta_{Z}+\delta_{n})$. In particular  it predicts the observed OHS
asymmetry  when changing the excitation polarization from $\sigma^{+}$ to $\sigma^{-}$ since
$\delta_{n}$ changes sign with respect to $\delta_{Z}$. Note the amplitude of this feedback
depends directly on the  finite nuclear spin depolarization time included in $\kappa$. Let us
now examine in more details the agreement of this model with our experimental results. For
$X^{+}$, the correlation time $\tau_{c}$ of the hyperfine interaction is given by its
 lifetime ($\sim$1~ns if radiatively limited) and the average electron spin is determined from the PL
circular polarization itself. The latter remains essentially constant around 70\%
[Fig.~\ref{OS&PL_vs_bias}] whereas the OHS increases continuously with the electric field
amplitude. This trend agrees thus well with Eq.~(\ref{OH}) when assuming a reduction of  trion
lifetime due to the competition with the field-induced electron escape. For $X^{0}$, in
addition to the anisotropic exchange term $\delta_{1}$ which averages to almost zero the
electron circular polarization $\langle S_{z}^{e}\rangle$ of bright states (see
Fig.~\ref{Fine-Structure}(c)), we should actually add to the spin splitting in Eq.~\ref{OH}
the direct exchange interaction ($\delta_{0}\approx$0.5~meV), since  the hyperfine hamiltonian
only couples bright to dark excitons. We thus expect  the nuclei polarization by $X^{0}$ to be
much more difficult than for trions. For $X^{-}$, the nuclear polarization dynamics is now
driven by the single electron left in the QD after optical recombination, whose lifetime
determines  $\tau_{c}$. Above 0.35V it is mainly limited by the capture time of a second
electron tunneling from the n-GaAs reservoir. The dependence observed in
Fig.~\ref{OS&PL_vs_bias}(b) showing a maximum of OHS at 0.45~V bias agrees qualitatively well
with a progressive reduction of $\tau_{c}$ leading first to the reduction of $\kappa$ and then
to the increase of $(\hbar/\tau_{c})^{2}$ in Eq.~\ref{OH}. However, since $\tau_{c}$,
$\rho_{c}$ and likely $T_{d}$ vary with the electric field it is not possible to fit the
results of Fig.~\ref{OS&PL_vs_bias}(b) without any additional assumptions. Therefore, to check
the validity of this model we have varied at fixed bias the average spin $\langle
S_{z}^{e}\rangle$ by rotating the quarter-wave plate which defines the excitation
polarization. The measurements of $\rho_{c}$ and $\delta_{n}$ are reported in
Fig.~\ref{OS&PL_vs_bias}(c) together with a theoretical fit. The good agreement, in particular
regarding the asymmetrical and non-linear dependence on $\rho_{c}$ gives a strong support to
our theoretical description. More intriguing, we couldn't observe the direct influence of the
hyperfine field fluctuations on the electron spin relaxation (i.e. on the circular
polarization of $X^{+}$)~\cite{PRB65-Merkulov2002} for the mere reason that the circular
polarization of $X^{+}$ was already quite strong in the 70$\--$80\% range in zero magnetic
field. This contrasts to Ref.~\cite{PRL94-Braun} where an ensemble of
p-doped QDs was studied. Further investigations are thus required to elucidate this   zero field result.\\
\indent In conclusion,  we have shown that a significative  nuclear polarization can be
induced by optically polarized trions  in a single InAs/GaAs quantum dot, with  a very
different signature according to the trion state $X^{+}$ or $X^{-}$. It demonstrates
unambiguously that the nuclear polarization results from the hyperfine interaction with an
unpaired electron  \textit{within} the quantum dot. Our results are discussed with a
theoretical model which provides a fine understanding of the nuclear polarization mechanism
and emphasizes the importance of   electron spin splitting on  its spin flip rate.


\begin{acknowledgments}
This work has been supported by  contract BoitQuant of \textit{Fonds National pour la Science}
and the European network of excellence SANDIE.
\end{acknowledgments}


\bibliographystyle{apsrev}


\begin{thebibliography}{24}
\bibitem[{Xpl()}]{XplusStar}
\bibinfo{note}{The weaker line denoted $X^{+^{\star}}$ is assigned to a hot
  trion because of its broad Lorentzian lineshape and its 5 meV redshift
  resulting of triplet configuration.}
\end{thebibliography}

\end{document}